%% file: RIS_Imaging_main.tex
\documentclass{article}
\usepackage{spconf}
\usepackage{algorithm,graphicx}
\usepackage[noend]{algpseudocode} 

\algrenewcommand\algorithmiccomment[1]{// {\itshape #1}}

\usepackage{url}
\usepackage{cite}
\usepackage[usenames]{color}
\usepackage{amsfonts}
\usepackage{acronym}
\usepackage{fancyhdr}
\usepackage{listings}%
\usepackage{arydshln}
\usepackage{amsmath,amssymb,amsthm}%
\usepackage{graphicx,psfrag,subcaption}
\usepackage{listings}%
\usepackage{arydshln}

\usepackage{bbm}
\input{macros}

\acrodef{ris}[RIS]{reconfigurable intelligent surface}
\acrodef{adc}[ADC]{analog-to-digital converter}

\title{Coded Aperture Radar Imaging Using \\ Reconfigurable Intelligent Surfaces}
\name{R.S. Prasobh Sankar and Sundeep Prabhakar Chepuri
	\thanks{This work is supported in part by the Next Generation Wireless Research and Standardization on 5G and Beyond project, MeitY, Government of India.
		}}
\address{Indian Institute of Science, Bangalore, India
	}

\begin{document}
	%
	\maketitle
	\begin{abstract} 
         In this paper, we focus on radar imaging using active sensing with a single transceiver and reconfigurable intelligent surface (RIS). RISs are arrays with tunable  passive phase shifter elements that can modify the propagation channel. The RIS reflects each transmit pulse with a different phase profile. We use compressive sensing to recover the radar scene from observations at the single-antenna receiver. We also provide a projected gradient descent algorithm to design the RIS phase shifts to obtain minimally coherent observations required for recovery. Through numerical simulations, we demonstrate that the proposed method recovers radar scenes with point scatterers and extended targets. 
         
	\end{abstract}

	\vspace*{-1mm}
	\maketitle
	

\section{Introduction}
Advances in sensors and computational methods have significantly enhanced our ability to detect targets and perform radar imaging accurately. To image a scene of interest, radar transmits a pulse and processes echoes reflected from  targets in the scene. Radar image includes a depth map or range and (azimuth and elevation) cross-range map of targets in the scene. It is not possible to reconstruct depth maps or resolve targets at different ranges, azimuth, and elevation angles using radars with a single antenna transceiver. Hence multiple input multiple outputs (MIMO) radars with rectangular or linear array geometries, each with a dedicated radio frequency (RF) chain, are typically used.

Although operating at higher frequencies (e.g., at millimeter wave bands) leads to a smaller form factor allowing many antenna elements to be packed in a small physical area, the RF front-end complexity (i.e., the number of \acp{adc} and other circuitry) also increases.  To reduce the RF front-end complexity due to multiple transceivers, several techniques are available, such as sparse arrays~\cite{moffet1968minimum,chen2008minimum}, (analog/digital) hybrid arrays~\cite{rajamaki2020hybrid}, or the usage of a low-resolution \ac{adc} for each antenna~\cite{ren2017one}, to name a few. Although sparse arrays, e.g., minimum redundancy arrays, have fewer antennas than uniformly spaced rectangular or linear arrays of equivalent aperture, both have the same target resolving ability. On the other hand, hybrid arrays also reduce the number of RF front ends by carefully combining signals at different antennas using analog phase shifters before digitizing them. An architecture with a single transceiver RF chain that employs analog beamforming to sense via \emph{coded beams} and image addition for reconstructing the scene has been studied in~\cite{lynch2018128,rajamaki2019analog}.
In a related context, assuming that the imaging scene is sparse, i.e., there are only a few point scatterers, compressive radar sensing techniques can reconstruct the imaging scene with much fewer observations~\cite{baraniuk2007compressive}. This work aims at reducing the RF front-end complexity by proposing an approach for radar imaging using a single-antenna transmitter and receiver aided with a \ac{ris}. 

\Ac{ris} is an emerging technology gaining significant interest in the communications and sensing domain because of its ability to modify the propagation channel favorably~\cite{renzo2020smart}. \Acp{ris} are rectangular arrays comprising a number of fully passive and remotely tunable phase shifters but without any other signal processing capability. These \ac{ris} phase shifters can be designed to control the  propagation channel, e.g., to beamform an incident signal to a desired direction or to create a virtual line-of-sight path between the transmitter and receiver. Due to the passive nature of \acp{ris}, they are more power efficient than multi-antenna transmitters with many RF chains. Although \acp{ris} have been primarily envisioned for wireless communications, it has also been studied for wireless sensing and localization applications~\cite{buzzi2022foundations,chepuri2023spm}. 

This paper proposes a new radar imaging approach with a single transceiver (hence a single RF chain), wherein the transmitter illuminates the \ac{ris} with a number of pulses. The \ac{ris} encodes each radiated pulse with different (deterministic or random) phase shifts leading to different phase profiles or \emph{coded apertures} and thereby introducing intentional delays. The reflected echoes from point scatterers corresponding to each reflection profile are processed digitally at the single-antenna receiver. We use sparse recovery techniques to resolve ambiguities in the measurements at the single sensor as the number of observations is much less than the number of pixels in the radar image scene. We also design the phase shifts of the \ac{ris} to obtain minimally coherent observations. Numerical experiments are provided to demonstrate the efficacy of the proposed approach. 

\section{System Model}
Consider a bistatic radar with one single-antenna transmitter and one single-antenna receiver at the spatial locations $\vl_{\rm tx}$ and $\vl_{\rm rx}$, respectively. The transmitter illuminates the \ac{ris}, which is a uniform rectangular array with $M$ tunable passive phase shifter elements. We assume that the transmitter and receiver are isolated from each other, the \ac{ris} is placed on the same side of the area of interest, and that the imaging scene is static.

Let $p(t)$ denote the time-domain pulse of width $\tau$ emitted by the transmitter. We transmit $N$ such pulses at a pulse repetition interval of $T_{p}$. The transmit pulse illuminates the $m$th \ac{ris} element located at the location $\vl_{{\rm ris},m}$. The transmitted signal at the $m$th \ac{ris} element undergoes an attenuation and delay as 
\[
q_m = \frac{\lambda\sqrt{\eta g(\alpha_m,\beta_m)}}{4\pi r_m} e^{-\jmath2\pi r_m/\lambda},
\]
where $\lambda = c/f_c$ with $c$ being the speed of the propagation medium and $f_c$ being the carrier frequency, $(r_m,\alpha_m,\beta_m)$ is the spherical coordinate of the $m$th \ac{ris} element with respect to the transmitter, $g(\alpha_m,\beta_m)$ is the antenna radiation pattern, and $\eta$ is the power efficiency of the transmit antenna. Each \ac{ris} element then phase shifts the incident signal at the $n$th transmission block by a phase $\phi_{m,n} \in [0,2\pi].$ Thus the time-domain signal from the $m$th \ac{ris} element at the $n$th time block is given by
\begin{equation}
    x_{m,n}(t) = q_m e^{-\jmath\phi_{m,n}}p(t)
       \label{eq:txpulse_ris}
\end{equation}
for $n=1,2,\ldots,N$. 
In the frequency domain, the signal in \eqref{eq:txpulse_ris}, denoted as $X_{m,n}(\omega)$, can be expressed as
\begin{equation*}
    X_{m,n}(\omega) = \int_{\mathbb{R}} x_{m,n}(t) e^{-\jmath\omega t} dt = q_m e^{-\jmath\phi_{m,n}} P(\omega),
\end{equation*}
where $P(\omega)$ is the transmit pulse in the frequency domain.

Consider a single point target located at $\vl_{\rm tar}$ with a reflectivity coefficient $r(\vl_{\rm tar})$, which is assumed to be flat across frequency. The received at the $n$th time block in the frequency domain can be expressed as
\begin{align}
    Y_{n}(\omega) &= \sum\limits_{m=1}^M r(\vl_{\rm tar}) X_{m,n}(\omega) e^{-\jmath\omega d_m/c} \notag\\
    &= r(\vl_{\rm tar}) P(\omega) \sum\limits_{m=1}^M q_m  e^{j( \phi_{m,n} + \omega d_m/c )},\notag
\end{align}
where $d_m = \|\vl_{{\rm ris},m} - \vl_{\rm tar} \| + \|\vl_{\rm tar} - \vl_{\rm rx}\|$ is the distance between the receiver and $m$th \ac{ris} element via the target. Here, the exponent term approximates the impulse response between the $m$th \ac{ris} element and receiver via the target.  When there are multiple point scatterers, the received signal contains a superposition of echoes from all the scatterers. Introducing different reflection patterns (aka phase shifts) of the \ac{ris} ${\boldsymbol \phi}_n = [\phi_{1,n},\phi_{2,n},\ldots,\phi_{M,n}]\rT$ at each time block behaves as a \emph{coded-aperture mask} that introduces intentional geometric delays. In this work, given such observations, we aim to reconstruct the radar image, which boils down to estimating the locations of the point scatterers. To this end, we assume that the radar image scene is sparse and use tools from compressive sensing in the next section.

\section{Compressive Radar Imaging}
Suppose there are $K$ pixels in the radar image scene and that the targets (or objects, each modeled with a scattering center) correspond to the pixel locations 
$\{\vl_{{\rm tar},k}, 1\leq k\leq K\}$ in the area of interest. The received signal is a superposition of echoes from all the $K$ target grid points and is modeled as 
\begin{align}
    Y_{n}(\omega) 
    &= \sum_{k=1}^Kr(\vl_{{\rm tar},k}) P(\omega) \sum\limits_{m=1}^M q_m e^{j( \phi_{m,n} + \omega d_{m,k}/c )}\notag\\
    &= \boldsymbol{\theta}_{n}\rT(\omega){\boldsymbol \Psi}(\omega)\vr
\end{align}
where $d_{m,k} = \|\vl_{{\rm ris},m} - \vl_{{\rm tar},k} \| + \|\vl_{{\rm tar},k} - \vl_{\rm rx}\|$ is the distance between the $m$th \ac{ris} element and the receiver via the $k$th target, 
$
\vr = [r(\vl_{{\rm tar},1}),\ldots,r(\vl_{{\rm tar},K})]\rT \in \mathbb{C}^{K},
$
and 
\[
{\boldsymbol \Psi}(\omega) =  \left[ {\boldsymbol \psi}(\omega,\vl_{{\rm tar},1}), \cdots, {\boldsymbol \psi}(\omega,\vl_{{\rm tar},K})
\right] \in \mathbb{C}^{M \times K}
\]
with the transmit waveform (aka sensing vector)
\[
\boldsymbol{\theta}_n(\omega) = [P(\omega)q_1e^{-\jmath\phi_{1,n}},\ldots,P(\omega)q_Me^{-\jmath\phi_{M,n}}]\rT
\in \mathbb{C}^{1 \times M}
\]
and the response vector of the \ac{ris} towards the $k$th target
\[{\boldsymbol \psi}(\omega,\vl_{{\rm tar},k}) = 
[e^{-\jmath \omega d_{1,k}/c }, \ldots, e^{-\jmath \omega d_{M,k}/c}]\rT \in \mathbb{C}^{M}.
\]
Thus the entries of $\vr$ will be non-zero whenever there are scatterers actually at those locations. We assume that there are very few scatters, say $T$, compared to the total number of pixels.

For $N$ different phase configurations of the \ac{ris} obtained by changing ${\boldsymbol \phi}_n$, we can describe the overall system as
\begin{equation}
\vy(\omega) = \boldsymbol{\Theta}(\omega) \mPsi(\omega) {\vr} = \mD \vr, 
\label{eq:datamodel}
\end{equation}
with
\[ 
\vy(\omega) = \begin{bmatrix}
Y_{1}(\omega) \\ \vdots \\
Y_{N}(\omega) 
\end{bmatrix} \text{  and  } \boldsymbol{\Theta}(\omega) =  \begin{bmatrix}
\btheta_{1}\rT(\omega) \\ \vdots \\
\btheta_{N}\rT(\omega)
\end{bmatrix} \in \mathbb{C}^{N \times M}.
\]
Although it is typical to discretize the frequency to a few points, for simplicity, we consider a single frequency and henceforth ignore the dependence of $\omega$ on the model in \eqref{eq:datamodel}. 
In the presence of noise, the above model becomes
\begin{equation} \label{eq:sparse:model1}
    \vy = \mD\vr + \vn,
\end{equation}
where $\vn \sim \cC\cN(\boldsymbol{0},\sigma^2{\bf I})$ is the receiver noise.

We aim to reconstruct the scattering scene from $\vy$ by estimating $\vr$. Let $\Vert \vr \Vert_0$ denote the $\ell_0$ norm or the number of non-zero entries in $\vr$. Then, to recover $\vr$ from $\vy$, we can search for a solution with exactly $T$ non-zero entries by solving
\begin{align}
    & \quad \underset{\vr}{\text{minimize}} \quad \Vert \vy - \mD\vr \Vert_2^2 \notag \\
    & \quad \text{subject to} \quad \Vert \vr \Vert_0 = T.
\end{align}
However, due to the presence of the sparsity constraint, the above optimization problem is non-convex and NP hard. Hence, we use a standard relaxation and replace the non-convex $\ell_0$ norm with a convex $\ell_1$ norm to obtain a convex optimization problem. The resulting convex optimization problem can be solved using any off-the-shelf solver.

\section{RIS Phase Profile Design}
A natural question to ask is, how to choose the \ac{ris} phase shifts in each time block? It is well-known that the sparse recovery performance improves by designing a measurement matrix such that its mutual coherence is the smallest~\cite{elad2007optimizedprojectionsparse,Abolghasemi2010optim_measurement_matrix}. For the measurement matrix $\mD = \boldsymbol{\Theta}\boldsymbol{\Psi}$, the mutual coherence is defined as the maximum value of the normalized inner product between its columns, i.e., 
\begin{equation}
    \mu(\mD) = \underset{i \neq j, 1 \leq i,j \leq K}{\text{max}}  \frac{\vert \vd_i\rH\vd_j \vert}{ \Vert \vd_i \Vert \Vert \vd_j \Vert } = \underset{i \neq j, 1 \leq i,j \leq K}{\text{max}} \vert [\mG]_{i,j} \vert,
\end{equation}
where $\mG = \Tilde{\mD}\rH \Tilde{\mD} \in \mbC^{K \times K}$ is the Gram matrix with $\tilde{\mD} \in \mbC^{N \times K}$
being the column-normalized version of ${\mD}$. Hence, reducing values of all the off-diagonal elements of the Gram matrix is sufficient to minimize the mutual coherence. Inspired by the sensing matrix design method in~\cite{Abolghasemi2010optim_measurement_matrix}, we next develop an algorithm to design the \ac{ris} phase shifts by making the Gram matrix close to the identity matrix to minimize all the off-diagonal entries and thereby reducing its mutual coherence.

Recall that $\boldsymbol{\Theta}$ depends on the \ac{ris} phase shifts and it can be alternatively expressed as
\begin{equation}
    \boldsymbol{\Theta} =  \boldsymbol{\Phi}\mQ,
\end{equation}
where $\mQ = P{\rm diag}\left(q_1,q_2,\cdots,q_M \right)$ is the diagonal matrix containing attenuation from the transmitter to the \ac{ris} with $P$ being the transmitted pulse in the frequency domain. Here, $\boldsymbol{\Phi}$ collects the \ac{ris} phase shifts as
\begin{equation}
    {\boldsymbol{\Phi}} = \begin{bmatrix}
        e^{-\jmath \phi_{1,1}}  & \ldots  & e^{-\jmath \phi_{M,1} } \\
        \vdots & \vdots & \vdots \\
        e^{-\jmath \phi_{N,1}} & \ldots & e^{-\jmath \phi_{N,M}}
    \end{bmatrix} \in \mbC^{N \times M},
\end{equation}
Thus the measurement matrix is 
\begin{equation}
    \mD = {\boldsymbol{\Phi}} \mQ \boldsymbol{\Psi}.
\end{equation}
The design of the compressed sensing measurement matrix reduces to the design of ${\boldsymbol{\Phi}}$ as $\mQ$ and ${\boldsymbol{\Psi}}$ are determined by the imaging scene.

The problem of designing the \ac{ris} phase shifts to minimize the mutual coherence can now be stated as
\begin{align}
    (\cP) \, & \quad  \underset{{\boldsymbol{\Phi}}}{ \text{minimize} } \quad J(\boldsymbol{{\Phi}}) = \Vert \boldsymbol{\Psi}\rH \mQ \rH \boldsymbol{{\Phi}}\rH \boldsymbol{{\Phi}} \mQ \boldsymbol{\Psi} - \mI  \Vert_F^2  \notag \\
    \, & \quad \text{subject to} \quad \vert [ {\boldsymbol{\Phi}} ]_{i,j} \vert = 1, \quad 1 \leq i,j \leq K, \label{eq:ris:phase}
\end{align}
where the unit modulus
constraint is due to the passive nature of the \ac{ris}. Here, we work with the unnormalized measurement matrix for simplicity.

To obtain the mutual coherence optimal phase shifts, we solve~($\cP$) in an iterative manner using projected gradient descent. 
Let ${\boldsymbol{\Phi}}_{k}$ denote the \ac{ris} phase shifts at iteration $k$. The update equations are given by
\begin{align} 
\label{eq:ris:update:grad}
    \tilde{\boldsymbol{\Phi}}_{k+1} &= {\boldsymbol{\Phi}}_k - \rho \nabla_{{\boldsymbol{\Phi}^*}} J({\boldsymbol{\Phi}}) \notag \\
    {\boldsymbol{\Phi}}_{k+1} &=  e^{{\jmath {\rm angle}( \tilde{\boldsymbol{\Phi}}_{k+1} )}},
\end{align}
where the gradient is given by
\begin{equation}
    \nabla_{{\boldsymbol{\Phi}^*}} J({\boldsymbol{\Phi}}) = {\boldsymbol{\Phi}}_k \mQ \boldsymbol{\Psi} ( \boldsymbol{\Psi}\rH \mQ \rH \boldsymbol{{\Phi}}_k\rH \boldsymbol{{\Phi}}_k \mQ \boldsymbol{\Psi} - \mI) \boldsymbol{\Psi}\rH \mQ\rH,
    \label{eq:gradient}
\end{equation}
where $(\cdot)^*$ denotes complex conjugation and $\rho$ is the step size which also includes the scaling term in  gradient~\eqref{eq:gradient}. The projection onto the unit circle ensures that each iterate ${\boldsymbol{\Phi}}_{k+1}$ has unit modulus entries.
We repeat~\eqref{eq:ris:update:grad} until convergence. This design procedure is summarized  as Algorithm~\ref{Alg_1}.
	\begin{algorithm}[!t] 
		\caption{RIS phase shift design}\label{Alg_1}
            {\bf Input:} $\boldsymbol{\Psi}$, $\mQ$, $\rho$ \\
            {\bf Output:} ${\boldsymbol{\Phi}}$ \\
		{\bf Initialization:} ${\boldsymbol{\Phi}}_0$ such that $|[{\boldsymbol{\Phi}}_0]_{i,j}|=1,\forall i,j$
		\begin{algorithmic}[1]
			\For{$k=1, 2, \cdots,\texttt{MaxIter}$}
			\State $\tilde{\boldsymbol{\Phi}}_{k+1} \gets {\boldsymbol{\Phi}}_k - \rho \nabla_{{\boldsymbol{\Phi}^*}} J({\boldsymbol{\Phi}})$ with $\nabla_{{\boldsymbol{\Phi}^*}}$ from \eqref{eq:gradient}
                \State ${\boldsymbol{\Phi}}_{k+1} \gets e^{{\jmath {\rm angle}( \tilde{\boldsymbol{\Phi}}_{k+1} )}}$
			\EndFor	
            \Return ${\boldsymbol{\Phi}}_{\texttt{MaxIter}}$
		\end{algorithmic}
	\end{algorithm}

\section{Experiments}

\begin{figure*}
\centering
\includegraphics[width=1.8\columnwidth]{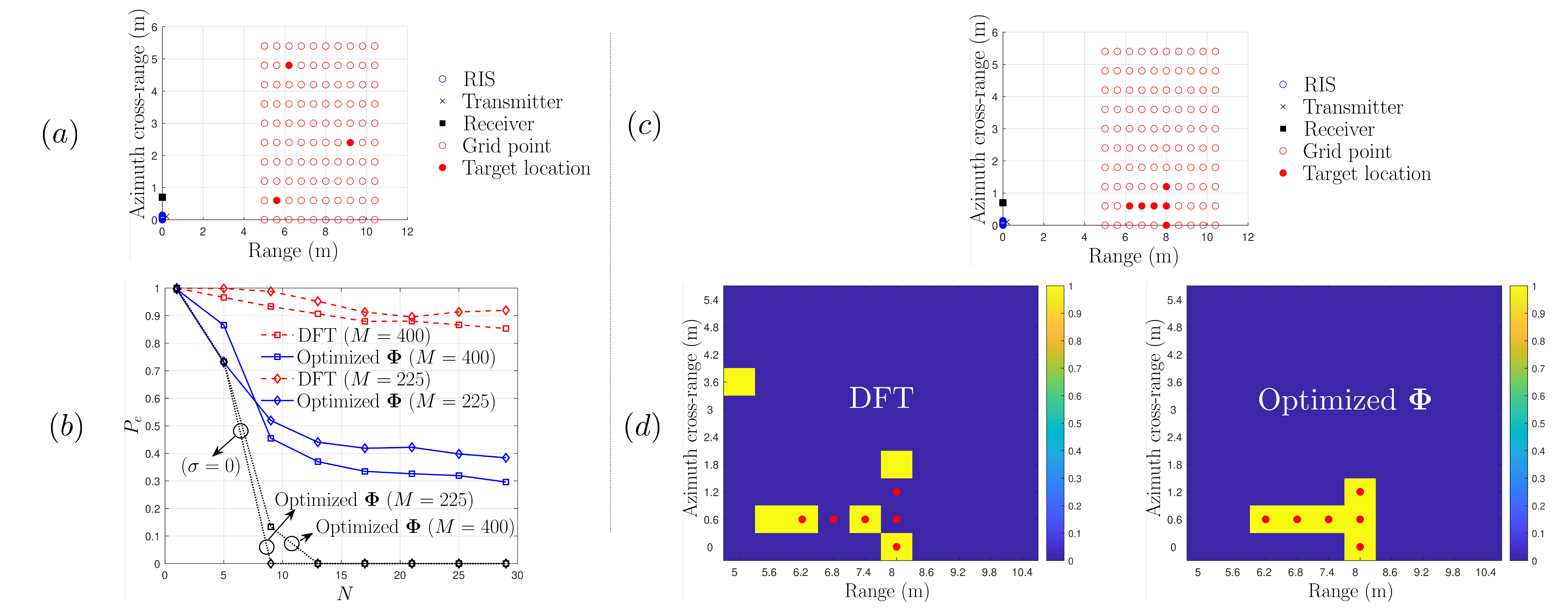}   
\caption{\small  (a) Sensing scenario with point scatterers. (b) Recovery performance. (c) Sensing scenario with an extended target. (d) Reconstructed scene.}
\label{fig:sim}
\vspace{-5mm}
\end{figure*}

This section presents several numerical experiments to demonstrate the proposed method. The transmitter and the receiver are located at $\vl_{\rm tx} = [0.2,0.1,0.1]$ m and $\vl_{\rm rx} = [0,0.7,0]$ m, respectively. The transmitter and the receiver are assumed to have a gain of $1$ towards the direction of \ac{ris} and a power efficiency $\eta=1$. The sensing system is assumed to operate at a carrier frequency of $f_c = 10$ GHz with a corresponding wavelength $\lambda=30$ cm and at a sampling rate of $0.5$ MHz. For the time domain pulse $p(t)$, we use a rectangular pulse of unit amplitude with a width of $\tau = 10 \mu$s. We use a measurement from a single frequency, i.e., $\omega = 2\pi 10^{10}$ rad/s and $\sigma = 0.01$.
The \ac{ris} is modeled as a uniform rectangular array with $M = 20\times 20$ elements with an inter-element spacing of $\lambda/4$. The RIS is located on the YZ plane with the bottom left corner at the origin $(0,0,0)$ m. To design the phase shifts, we use a step size of $\rho=0.01$ and \texttt{MaxIter}$=1000$. Next to the designed phase shift matrix according to Algorithm~\ref{Alg_1}, we also use the first $N$ rows of an $M \times M$ DFT matrix as the phase shift matrix for comparison.

We  assume that the true targets lie on the target grid (i.e., the image pixel). To form the dictionary $\boldsymbol{\Psi}$, we discretize the target scene into a rectangular grid of $K=100$ points with $10$ grid points along the range and $10$ grid points along the (azimuth) cross-range, with each grid point separated by $20\lambda \approx 60$ cm. We evaluate the performance of the proposed method in terms of the probability of error in support recovery $P_{\rm e}$. Specifically, let $i_{{\rm tar},k}$ and $\hat{i}_{{\rm tar},k}$ denote the true and estimated grid indices of the $k$th target, respectively.  Let $T$ be the number of targets. We define $P_{\rm e}$ as 
\begin{equation}
    P_{\rm e} = \mathbb{E}\left[ \frac{1}{T} \sum_{k=1}^T  \mathbbm{1}(i_{{\rm tar},k} \neq \hat{i}_{{\rm tar},k}) \right],
\end{equation}
where the expectation is with respect to different noise realizations and  $\mathbbm{1}(\cdot)$ is an indicator function with $\mathbbm{1}(\text{TRUE})=1$ and $\mathbbm{1}(\text{FALSE})=0$. In this work, we compute the expectation empirically using $500$ independent noise realizations.

We begin by considering a scenario where each target is modeled as a point scatterer. Specifically, we consider $3$ point targets at locations $(5.6,0.6,0)$ m, $(9.2,2.4,0)$ m, and $(6.2,4.8,0)$ m as shown in Fig.~\ref{fig:sim}(a). We present the impact of the number of measurements $N$ on $P_{\rm e}$ for different values of $M$ in Fig.~\ref{fig:sim}(b). For reference, we have also presented the performance of the proposed scheme in a noiseless setting. The proposed scheme is significantly better than the DFT matrix-based \ac{ris} phase shifts.  With $M=400$, the proposed scheme succeeds in estimating the correct support for more than $60\%$ with as few as  $12$
measurements. On the other hand, for \texttt{DFT}, the correct support is only estimated in about $10\%$ for $N=12$, clearly demonstrating the necessity to appropriately design the \ac{ris} phase shifts.
As expected, with an increase in $N$, the sparse recovery algorithms perform better and lead to a smaller $P_{\rm e}$ with the proposed design. For large values of $N$, we observe that the performance saturates for the noisy scenario. This is because of the inevitable error floor arising from additive noise at the receiver. On the other hand, for a noiseless setting ($\sigma=0$), the sparse recovery is perfect once we acquire the required minimum number of measurements. We also observe that the sensing performance, in general, increases with an increase in the size of \ac{ris}.

Next, we consider a more challenging scenario with extended targets.  For this setting, we use $\sigma = 0$. We aim to sense an object, which we model as a collection of several point scatterers. Specifically, we sense an object having the shape of the English alphabet ``T" as illustrated in Fig.~\ref{fig:sim}(c). We use a grid size of $K=100$ and $N=20$ measurements to reconstruct the scene. The reconstructed scene is shown in Fig.~\ref{fig:sim}(d). The proposed scheme reconstructs the target scene reasonably well despite having only a single RF chain on both the transmitter and the receiver, clearly demonstrating the advantages of using an \ac{ris} for radar imaging.

\vspace{-4mm}

\section{Conclusions}
\vspace{-3mm}

In this paper, we considered the problem of radar imaging using active sensing in a system with a single antenna transmitter, a single antenna receiver, and an \ac{ris}. Specifically, we used \ac{ris} as a coded aperture mask to introduce intentional geometric delays through different phase profiles. We developed a sparse sensing-based framework for radar imaging with \ac{ris} and presented an iterative algorithm to design the \ac{ris} phase shifts to obtain minimally coherent measurements at the receiver. Through numerical simulations, we demonstrated that the proposed method offers superior sensing performance compared to a system with arbitrary \ac{ris} phase shifts for both point and extended targets.
	
\pagebreak
\bibliographystyle{IEEEtran}
\bibliography{IEEEabrv,refs}
	
\end{document}

%% file: macros.tex
\def\cast{{
   \mathord{
      \hbox to 0em{
         \ooalign{
	   \smash{\hbox{$\ast$}}\crcr
	   \smash{\hskip-1pt\Large\hbox{$\circ$}} }
	 \hidewidth}
      \phantom{\bigcirc}
} }}

\def\bm#1{\mbox{\boldmath $#1$}}

\newcommand{\mPsi}{\mbox{$\bm \Psi$}}

\newcommand{\rH}{^{ \raisebox{1pt}{$\rm \scriptscriptstyle H$}}}
\newcommand{\rT}{^{ \raisebox{1.2pt}{$\rm \scriptstyle T$}}}

\newcommand{\bds}{\begin {itemize}}
\newcommand{\eds}{\end {itemize}}
\newcommand{\bdf}{\begin{definition}}
\newcommand{\blm}{\begin{lemma}}
\newcommand{\edf}{\end{definition}}
\newcommand{\elm}{\end{lemma}}
\newcommand{\bthm}{\begin{theorem}}
\newcommand{\ethm}{\end{theorem}}
\newcommand{\bprp}{\begin{prop}}
\newcommand{\eprp}{\end{prop}}
\newcommand{\bcl}{\begin{claim}}
\newcommand{\ecl}{\end{claim}}
\newcommand{\bcr}{\begin{coro}}
\newcommand{\ecr}{\end{coro}}
\newcommand{\bquest}{\begin{question}}
\newcommand{\equest}{\end{question}}

\newcommand{\larrow}{{\larrow}}

\newcommand{\argmin}{\ensuremath{\mathrm{arg}\min}}
\newcommand{\argmax}{\ensuremath{\mathrm{arg}\max}}

\newcommand{\cC}{{\ensuremath{\mathcal{C}}}}

\newcommand{\cN}{{\ensuremath{\mathcal{N}}}}

\newcommand{\cP}{{\ensuremath{\mathcal{P}}}}

\newcommand{\vd}{{\ensuremath{{\mathbf{d}}}}}

\newcommand{\vl}{{\ensuremath{{\mathbf{l}}}}}

\newcommand{\vn}{{\ensuremath{{\mathbf{n}}}}}

\newcommand{\vr}{{\ensuremath{{\mathbf{r}}}}}

\newcommand{\vy}{{\ensuremath{{\mathbf{y}}}}}

\newcommand{\mD}{{\ensuremath{\mathbf{D}}}}

\newcommand{\mG}{{\ensuremath{\mathbf{G}}}}

\newcommand{\mI}{{\ensuremath{\mathbf{I}}}}

\newcommand{\mQ}{{\ensuremath{\mathbf{Q}}}}

\usepackage{latexsym}

\def\IC{\mathbb C}
\def\IN{\mathbb N}
\def\IZ{\mathbb Z}
\def\IR{\mathbb R}

\newcommand{\mbC}{{\ensuremath{\mathbb{C}}}}

\def\shat{^{\mathchoice{}{}%
 {\,\,\smash{\hbox{\lower4pt\hbox{$\widehat{\null}$}}}}%
 {\,\smash{\hbox{\lower3pt\hbox{$\hat{\null}$}}}}}}

\def\bSigma{{
      \ooalign{
      \smash{\hskip.4pt\raise.4pt\hbox{$\Sigma$}}\vphantom{}\crcr
      \smash{\hskip.7pt\raise.6pt\hbox{$\Sigma$}}\vphantom{}\crcr
      \smash{\hbox{$\Sigma$}}\vphantom{$\Sigma$}}
      \vphantom{\hbox{$\Sigma$}}
      }}
\def\bTheta{{
      \ooalign{
      \smash{\hskip.5pt\raise.5pt\hbox{$\Theta$}}\vphantom{}\crcr
      \smash{\hskip.0pt\raise.1pt\hbox{$\Theta$}}\vphantom{}\crcr
      \smash{\hbox{$\Theta$}}\vphantom{$\Theta$}}
      \vphantom{\hbox{$\Theta$}}
      }}
\def\bDelta{{
      \ooalign{
      \smash{\hskip.4pt\raise.4pt\hbox{$\Delta$}}\vphantom{}\crcr
      \smash{\hskip.7pt\raise.6pt\hbox{$\Delta$}}\vphantom{}\crcr
      \smash{\hbox{$\Delta$}}\vphantom{$\Delta$}}
      \vphantom{\hbox{$\Delta$}}
      }}
\def\bLambda{{
      \ooalign{
      \smash{\hskip.5pt\raise.5pt\hbox{$\Lambda$}}\vphantom{}\crcr
      \smash{\hskip.0pt\raise.1pt\hbox{$\Lambda$}}\vphantom{}\crcr
      \smash{\hbox{$\Lambda$}}\vphantom{$\Lambda$}}
      \vphantom{\hbox{$\Lambda$}}
      }}

\makeatletter

\def\bordermatrix#1{\begingroup \m@th
  \@tempdima 8.75\p@
  \setbox\z@\vbox{%
    \def\cr{\crcr\noalign{\kern2\p@\global\let\cr\endline}}%
    \ialign{$##$\hfil\kern2\p@\kern\@tempdima&\thinspace\hfil$##$\hfil
      &&\quad\hfil$##$\hfil\crcr
      \omit\strut\hfil\crcr\noalign{\kern-\baselineskip}%
      #1\crcr\omit\strut\cr}}%
  \setbox\tw@\vbox{\unvcopy\z@\global\setbox\@ne\lastbox}%
  \setbox\tw@\hbox{\unhbox\@ne\unskip\global\setbox\@ne\lastbox}%
  \setbox\tw@\hbox{$\kern\wd\@ne\kern-\@tempdima\left[\kern-\wd\@ne
    \global\setbox\@ne\vbox{\box\@ne\kern2\p@}%
    \vcenter{\kern-\ht\@ne\unvbox\z@\kern-\baselineskip}\,\right]$}%
  \null\;\vbox{\kern\ht\@ne\box\tw@}\endgroup}
\makeatother

\makeatletter
\def\argmin{\mathop{\operator@font arg\,min}}
\def\argmax{\mathop{\operator@font arg\,max}}
\makeatother

\def\bm#1{\mbox{\boldmath $#1$}}

\newcommand{\bbeta}{{\bm\beta}}

\newcommand{\btheta}{{\bm\theta}}

\newcommand{\bea}{\begin{array}}
\newcommand{\ena}{\end{array}}
\newcommand{\beq}{\begin{equation}}
\newcommand{\enq}{\end{equation}}

\newcommand{\beqa}{\begin{eqnarray}}
\newcommand{\enqa}{\end{eqnarray}}

\newcommand{\beqan}{\begin{eqnarray*}}
\newcommand{\enqan}{\end{eqnarray*}}

\newcommand{\AL}{\begin{enumerate}}
\newcommand{\ALE}{\end{enumerate}}

\def\addots{\mathinner{
    \mkern1mu\raise0pt\vbox{\kern7pt\hbox{.}}
    \mkern2mu\raise4pt\hbox{.}
    \mkern2mu\raise7pt\hbox{.}
    \mkern1mu}}

\def\sddots{\mathinner{
    \mkern.8mu\raise7pt\hbox{.}
    \mkern.8mu\raise4pt\hbox{.}
    \mkern.8mu\raise0pt\vbox{\kern7pt\hbox{.}}
    \mkern1mu}}

\def\saddots{\mathinner{
    \mkern.2mu\raise0pt\vbox{\kern7pt\hbox{.}}
    \mkern.2mu\raise4pt\hbox{.}
    \mkern.2mu\raise7pt\hbox{.}
    \mkern1mu}}

\def\sqplus{\mathbin{
	{\ooalign{\hfil\raise.3ex\hbox{\scriptsize
	+}\hfil\crcr\mathhexbox274\crcr\mathhexbox275}}
	}} 
\def\sqminus{\mathbin{
	{\ooalign{\hfil\raise.3ex\hbox{\scriptsize
	--}\hfil\crcr\mathhexbox274\crcr\mathhexbox275}}
	}}

\def\IC{{
   \mathord{
      \hbox to 0em{
	 \hskip-4pt
         \ooalign{
	   \smash{\hskip1.9pt\raise2.6pt\hbox{$\scriptscriptstyle |$}}\crcr
	   \smash{\hbox{\rm\sf C}} }
	 \hidewidth}
      \phantom{\hbox{\rm\sf C}}
} }}
\def\IN{
    {\ooalign{
   \smash{\hskip2.2pt\raise1.5pt\hbox{$\scriptscriptstyle |$}}\vphantom{}\crcr
   \hbox{\sf N}
	}}
	} 
\def\IZ{
    {\ooalign{
   \smash{\hskip1.9pt\raise0pt\hbox{$\sf Z$}}\vphantom{}\crcr
   \hbox{\sf Z}
	}}
	} 
\def\IR{
    {\ooalign{
   \smash{\hskip2.2pt\raise1.5pt\hbox{$\scriptscriptstyle |$}}\vphantom{}\crcr
   \smash{\hskip2.2pt\raise3.3pt\hbox{$\scriptscriptstyle |$}}\vphantom{}\crcr
   \hbox{\sf R}
	}}
	} 

\DeclareMathAlphabet{\mathcmb}{OT1}{cmr}{b}{n}

\def\bSigma{\ensuremath{\mathcmb{\Sigma}}}
\def\bLambda{\ensuremath{\mathcmb{\Lambda}}}

\def\bTheta{\ensuremath{\mathcmb{\Theta}}}

\newcommand{\SI}{\begin{indlist}}
\newcommand{\EI}{\end{indlist}}

\newcommand{\DL}{\begin{dashlist}}
\newcommand{\DLE}{\end{dashlist}}

\makeatletter
\def\setboxz@h{\setbox\z@\hbox}
\def\wdz@{\wd\z@}
\def\boxz@{\box\z@}
\def\underset#1#2{\binrel@{#2}%
  \binrel@@{\mathop{\kern\z@#2}\limits_{#1}}}
\def\binrel@#1{\begingroup
  \setboxz@h{\thinmuskip0mu
    \medmuskip\m@ne mu\thickmuskip\@ne mu
    \setbox\tw@\hbox{$#1\m@th$}\kern-\wd\tw@
    ${}#1{}\m@th$}%
  \edef\@tempa{\endgroup\let\noexpand\binrel@@
    \ifdim\wdz@<\z@ \mathbin
    \else\ifdim\wdz@>\z@ \mathrel
    \else \relax\fi\fi}%
  \@tempa
}
\let\binrel@@\relax%
\makeatother
